\documentclass[reqno]{amsart}
\usepackage[utf8]{inputenc}
\usepackage[T1]{fontenc}								
\usepackage{graphicx}									
\usepackage{subfig}										
\usepackage{color}										
\usepackage{lmodern}									
\usepackage{amssymb}									
\usepackage{amsthm}										
\usepackage{listliketab}								
\usepackage{multirow}									
\usepackage{bbm}										
\usepackage{float}
\usepackage[font=small]{caption}
\usepackage{amsmath}

\usepackage{setspace}
\AtBeginDocument{%
  \addtolength\abovedisplayskip{0.5\baselineskip}%
  \addtolength\belowdisplayskip{0.5\baselineskip}%
}

\title[The Pattern Recognition of Probability Distributions of amino acids]
{The Pattern Recognition of Probability Distributions of amino acids in protein families}

\author{R.P.~Mondaini, S.C.~de Albuquerque Neto}  
\address{Federal University of Rio de Janeiro\\
Centre of Technology, COPPE\\
Rio de Janeiro - RJ, Brazil}
\email[R.~Mondaini]{Rubem.Mondaini@ufrj.br}

\begin{document}

\begin{abstract}
A Pattern Recognition of a Probability Distribution of amino acids is obtained for selected families of proteins. The
mathematical model is derived from a theory of protein families formation which is derived from application of a Pauli's master
equation method.
\end{abstract}

\maketitle

\section{Introduction}
The formation and evolution of a protein family is a problem of the same importance as that of protein folding and unfolding. We also
think that the last problem will be solved or at least treated on a more general perspective, by concentrating the theoretical
research on the joint formation and evolution of the whole set of proteins of each protein family. A probabilistic analysis of a
model for protein family formation is then most welcome which is able to unveil the specific nature of this protein family formation
process (PFFP) and the consequent folding/unfolding process. A pictorial representation of the PFFP to be seen as a game for the
upsurge of life and its homeostasis can be introduced by thinking on $\mathbf{n}$ consecutive trials of $\mathbf{m}$ icosahedra, each
face of them corresponding to a different amino acid. This sort of ideas has been already introduced on the scientific literature of
the Entropy Maximization Principle \cite{jaynes,harte}. Our desiderata is then to translate the information contained on biological
almanacs (protein databases) in terms of random variables in order to model a dynamics for describing the folding and unfolding of
proteins, This means that we think on PFFP instead of the evolution of a single protein as the key to understand the protein dynamics.

In section 2, we introduce a description of the sample space of probability to be used in the calculations of the present
contribution \cite{mondaini1,mondaini2}. We have also made a digression on generalized proposals of joint probabilities which should
be used on future generalizations of the present work. In section 3, a Poisson statistical receipt is derived from applications of a
Master Equation method in order to derive one adequate probability distribution of our PFFP problem \cite{vanKampen,bialek} in section
4. In section 5, we describe the distribution of amino acids in protein families and the pattern recognition through level curves of
the probability distribution. We also determine the domain of the variables for the present probabilistic model. In section 6, we
introduce a scheme in terms of graphical cartesian representations of level curves with selected protein families from the Pfam
database \cite{finn1,punta,finn2,finn3}. We also investigate the possibility of using this representation to justify the
classification of protein families into clans. A final section of concluding remarks will comment on the introduction of alternative
candidates to usual probability distributions, like the use of joint probabilities and the study of Non-additive entropy measures.
All these proposals which aim to improve the pattern recognition method will appear in forthcoming publications.

\section{The organization of the sample space of probability}
The fundamental idea of the Protein Family Formation Process as introduced in section 1, is of a dynamical game played by nature. The
amino acids which are necessary to participate in the formation of a protein family are obtained from a ``universal ribossome
deposit'' and the process will consist in the distribution on shelves of a bookshelf. In the first stage, $\mathbf{m}$ amino acids
are distributed on $\mathbf{m}$ shelves of the first bookshelf. On a second stage, the $\mathbf{m}$ shelves of the second bookshelf
should be fulfilled by the amino acids transferred from the first bookshelf and so on. After choosing a database, in order to
implement all the future calculations with random variables, we select an array of $\mathbf{m}$ protein domains (rows). There are
$n_1$ amino acids on the first column, $n_2$ in the second one, \ldots $n_m$ in the $m^{\mathrm{th}}$ column. We then select the
($m \, \mathrm{x} \, n$) block where
\begin{equation}
 n = \min (n_1, n_2, \ldots, n_m)
\end{equation}

Another way of introducing a ($m \, \mathrm{x} \, n$) block is to specify the number $\mathbf{n}$ of columns a priori and to delete
all proteins such that $n_r<n$, $r = 1,2, \ldots, m$ and to also delete the last $(n_r - n)$ amino acids on the remaining proteins.
In fig.1 below we present an example of a $(m\, \mathrm{x} \,n)$ block, which is organized according to this second basic
procedure. There is at least one block associated to a protein family.

\begin{figure}[!hb]
 \centering
 \includegraphics[width=1\linewidth]{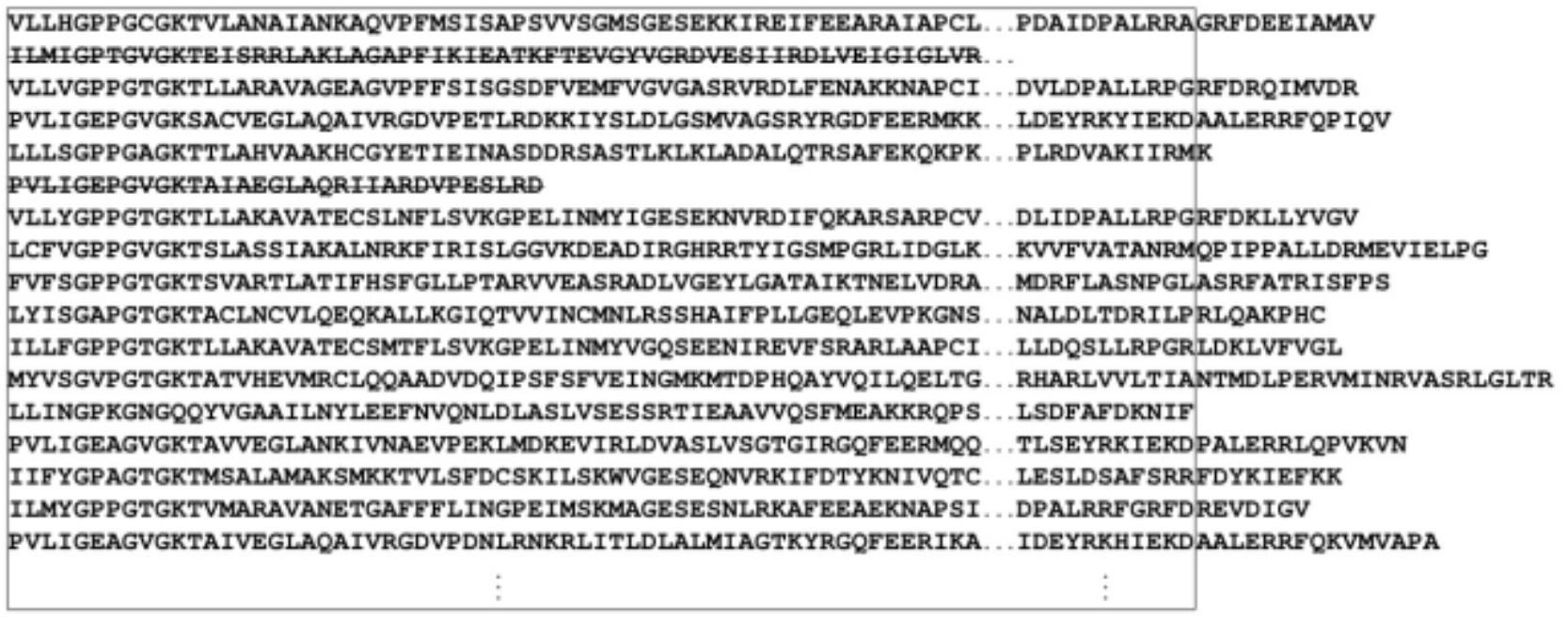}
 \caption{A block of $(m = 100) \, \mathrm{x} \, (n = 200)$ amino acids from Pfam database.}
\end{figure}

Let $p_j(a)$ be the probability of occurrence of the amino acid and $a=$ A, C, D, E, F, G, H, I, K, L, M, N, P, Q, R, S, T, V, W, Y
in the $j$-th column,
\begin{equation}
 p_j(a) = \frac{n_j(a)}{m}
\end{equation}
where $n_j(a)$ is the number of occurrences of the $a$-amino acid in the $j$-th column of the $(m \, \mathrm{x} \, n)$
block.

We have
\begin{equation}
 \sum_a n_j(a) = m \, , \quad \forall j
\end{equation}
and
\begin{equation}
 \sum_a p_j(a) = 1 \, , \quad \forall j
\end{equation}

The probabilities $p_j(a)$ will be considered as the components of a $20$-component vector, $\overrightarrow{p_j}=1,2,\ldots,n$.
These $n$ vectors will be random variables w.r.t. the probability distribution already defined in eq.$(2)$.

We can also introduce the probability distribution corresponding to the occurrence of amino acids $a$, $b$ in columns $j$, $k$,
respectively. We can write,
\begin{equation}
 P_{jk}(a,b) = \frac{n_{jk}(a,b)}{m}
\end{equation}
with $a, b =$ A, C, D, E, F, G, H, I, K, L, M, N, P, Q, R, S, T, V, W, Y.

This is the joint probability distribution and it can be understood in terms of the conditional probability $P_{jk}(a|b)$ as
\begin{equation}
 P_{jk}(a,b) = P_{jk}(a|b)p_k(b)
\end{equation}

From Bayes' law [11] we can write
\begin{equation}
 P_{jk}(a,b) = P_{jk}(a|b)p_k(b) = P_{kj}(b|a)pj(a) = P_{kj}(b,a)
\end{equation}
Analogously to eqs.$(3)$,$(4)$, we can write
\begin{equation}
 \sum_a \sum_b n_{jk}(a,b) = m , \quad \forall j,k , \quad j<k
\end{equation}
and
\begin{equation}
 \sum_a \sum_b P_{jk}(a,b) = \sum_a p_j(a) = 1 , \quad \forall j,k , \quad j<k
\end{equation}
We can take on eq.$(8)$, $(9)$:
\begin{equation*}
 j = 1, 2, \ldots, (n-1), k = (j+1), (j+2), \ldots, n.
\end{equation*}
These random variables will be arranged as $\binom{n}{2} = \frac{n(n-1)}{2}$ square matrices of $20^{\mathrm{th}}$ order.

A straightforward generalization to a multiplet of $\mathbf{s}$ amino acids, which occur on $\mathbf{s}$ ordered columns, can be
done by introducing joint probabilities such as,
\begin{equation}
 P_{j_1 j_2 \ldots j_s}(a_1,a_2,\ldots,a_s) = \frac{n_{j_1 j_2 \ldots j_s}(a_1,a_2,\ldots,a_s)}{m}
\end{equation}
where
\begin{equation}
 a_1,a_2,\ldots,a_s = A, C, D, E, F, G, H, I, K, L, M, N, P, Q, R, S, T, V, W, Y
\end{equation}
and
\begin{equation*}
 j_1 < j_2 < \ldots < j_s
\end{equation*}
with
\begin{align*}
 j_1 &= 1, 2, \ldots , (n-s+1) \\
 j_2 &= (j_1 + 1), (j_1 + 2), \ldots , (n-s+2) \\
 \vdots & \hspace{1cm} \vdots \hspace{2cm} \vdots \\
 j_s &= (j_{s-1} + 1), (j_{s-1} + 2), \ldots , n 
\end{align*}

These random variables $P_{j_1 j_2 \ldots j_s}(a_1,a_2,\ldots,a_s)$ are $\binom{n}{s} = \frac{n!}{s!(n-s)!}$ objects of $(20)^s$
components each. Analogously to eqs.$(8)$, $(9)$, we can now write,
\begin{align}
\sum_{a_1} \sum_{a_2} \ldots \sum_{a_s} n_{j_1 j_2 \ldots j_s}(a_1,a_2,\ldots,a_s) = m \, , \quad \forall j_1, j_2, \ldots, j_s \\
 j_1 < j_2 < \ldots < j_s \nonumber
\end{align}
and
\begin{align}
\sum_{a_1} \sum_{a_2} \ldots \sum_{a_s}& P_{j_1 j_2 \ldots j_s}(a_1,a_2,\ldots,a_s) = \ldots = \sum_{a_1} \sum_{a_2}
P_{j_1 j_2}(a_1,a_2) \nonumber \\
=& \sum_{a_1} p_{j_1}(a_1) = 1 \, , \quad \forall j_1, j_2, \ldots, j_s \, , \, j_1 < j_2 < \ldots < j_s \label{sumjointsprob}
\end{align}

Eqs.$(10)$-$(13)$ will be reserved for future developments. In the present work, we restrict all calculations to simple
probabilities given by eq.$(2)$-$(4)$.

\section{The Master Equation for Probability Evolution}
The temporal evolution of random variables such as the probabilities of occurrence introduced above can be modelled through a master
equation approach \cite{vanKampen,bialek}.

Let $p\left(n_j\big(t(a)\big)\right)$ be the probability of occurrences of the amino acids $a$ in the $j$-th column of the
($m \, \mathrm{x} \, n$) block at time $t(a)$. The probability of observing the same amino acid at the $j$-th column after an
interval of time $\Delta t$ is given by:
\begin{equation}
 \resizebox{.9\hsize}{!}{$p\left(n_j\big(t(a)+\Delta t\big)\right) = \sigma\left(t(a)\right)\Delta t \,\, p\left(n_{j-1}
 \big(t(a)\big)\right) + \left(1 + \sigma\big(t(a)\big)\right)\Delta t \,\, p\left(n_j \big(t(a)\big) \right)$}
\end{equation}
where $\sigma\left(t(a)\right)$ is the transition probability per unit time between columns $j-1$ and $j$.

We now imagine that there is a column $j=0$ -- ``the universal ribosome deposit'' where all amino acids are present at time
$t_0(a)$, as
\begin{equation}
 p\left(n_0\big(t_0(a)\big)\right) = 1 \, , \quad \forall a
\end{equation}

This also means that at the initial time $t_0(a)$ no amino acid has been received by the columns $j \neq 0$ -- the shelves of amino
acid bookshelf (the protein family), as
\begin{equation}
 p\left(n_j\big(t_0(a)\big)\right) = 0 \, , \quad \forall j \neq 0 \, , \quad \forall a
\end{equation}

After taking the limit $\Delta t \to 0$ on eq.$(14)$, we get:
\begin{equation}
 \frac{\partial p\left(n_j\big(t(a)\big)\right)}{\partial t(a)} = \sigma\left(t(a)\right)\left(p\left(n_{j-1}\big(t(a)\big)\right)
 - p\left(n_j\big(t(a)\big)\right) \right) \, , \quad j \neq 0
\end{equation}
We also have, for $j = 0$,
\begin{equation}
 \frac{\partial p\left(n_0\big(t(a)\big)\right)}{\partial t(a)} = -\sigma\left(t(a)\right) \, p\left(n_0\big(t(a)\big)\right)
\end{equation}

From eqs.$(15)$ and $(18)$, we have,
\begin{equation}
 p\left(n_0\big(t(a)\big)\right) = e^{-v\big(t(a)\big)}
\end{equation}
where
\begin{equation}
 v\big(t(a)\big) = \int_{t_0(a)}^{t(a)} \sigma \big(t'(a)\big) \mathrm{d}t'(a)
\end{equation}

From eqs.$(16)$ and $(17)$ we can write for $j=1$
\begin{equation}
 p\left(n_1\big(t(a)\big)\right) = e^{-v\left(t(a)\right)}v\left(t(a)\right)
\end{equation}
We also have for $j=2$,
\begin{equation}
 p\left(n_2\big(t(a)\big)\right) = e^{-v\left(t(a)\right)} \left( \int_{t_0(a)}^{t(a)}
 \frac{\mathrm{d}v\left(t'(a)\right)}{\mathrm{d}t'(a)} v\left(t'(a)\right) \mathrm{d}t'(a) \right)
\end{equation}
From eqs.$(15)$, $(19)$, we have
\begin{equation}
 v\left(t_0(a)\right) = 0
\end{equation}
Eq.$(22)$ will turn into:
\begin{equation}
 p\left(n_2\big(t(a)\big)\right) = e^{-v\left(t(a)\right)}\frac{v^2\left(t(a)\right)}{2}
\end{equation}
By finite induction on $j$, we can write the Poisson distribution:
\begin{equation}
 p\left(n_j\big(t(a)\big)\right) = e^{-v\left(t(a)\right)}\frac{v^j\left(t(a)\right)}{j!} \, , \quad \forall j \, , \quad \forall a
\end{equation}

\section{The Distribution of amino acids as a Marginal Probability Distribution}
We now introduce the marginal probability distributions \cite{deGroot} associated to the Poisson process given by eq.$(25)$. We have:
\begin{equation}
 p_j\big(t(a)\big) = \int_{t_0(a)}^{t(a)} p \left(n_j\big(t'(a)\big)\right) \mathrm{d}t'(a)
\end{equation}
From eqs.$(25)$, $(26)$, we can write:
\begin{equation}
 p_j\big(t(a)\big) = \frac{(-1)^j}{j!} \lim_{\alpha \to 1} \frac{\partial^j}{\partial\alpha^j} \int_{t_0(a)}^{t(a)}
 e^{-\alpha v\left(t'(a)\right)} \mathrm{d}t'(a)
\end{equation}
where $\alpha$ is an auxiliary parameter.

In the present work, we make the assumption that $\sigma\left(t(a)\right) \equiv \sigma(a)$, which leads to a linear approximation
for $v\left(t(a)\right)$ through eq.$(20)$, or,
\begin{equation}
 v\left(t(a)\right) = \sigma(a)\left(t(a)-t_0(a)\right)
\end{equation}
We then have from eqs.$(27)$,$(28)$,
\begin{equation}
 p_j\big(t(a)\big) = \frac{(-1)^{j-1}}{j!\sigma(a)} \lim_{\alpha \to 1} \frac{\partial^j}{\partial\alpha^j}
 \left(\frac{e^{-\alpha\sigma(a)\left(t(a)-t_0(a)\right)} -1}{\alpha}\right)
\end{equation}
Let us write now $t_j(a)$ as the time in which the $a$-amino acid is seen to occur at the $j$-th column of the $(m
\, \mathrm{x} \,n)$ block. 

We write,
\begin{equation}
 t_j(a) = t_0(a) + j \Delta(a)
\end{equation}
where $\Delta(a)$ is the time interval for the transition of the amino acid between consecutive columns,
\begin{equation}
 \Delta(a) = t_j(a) - t_{j-1}(a)\, , \, j=1,2,\ldots,n
\end{equation}

The marginal probability distribution function of eq.$(26)$ should be considered as a two-variable distribution:
\begin{equation}
 p_j\big(t(a)\big) \equiv p_j\big(\sigma_j(a), \Delta(a)\big) = \frac{(-1)^{j-1}}{j! \, \sigma(a)} \lim_{\alpha \to 1}
 Q_j\big(\alpha\,; \sigma_j(a) \Delta(a)\big)
\end{equation}
where
\begin{equation}
 \resizebox{.9\hsize}{!}{$Q_j\big(\alpha\,; \sigma_j(a) \Delta(a)\big) = (-1)^j \, j! \left(e^{-\alpha j \, \sigma_j(a)\Delta(a)}
 \left(\alpha^{-(j+1)} \sum\limits_{m=1}^{j} \frac{\big(j \, \sigma_j(a)\Delta(a)\big)^m \alpha^{-(j-m+1)}}{m!} \right) -1\right)$}
\end{equation}
we can then write,
\begin{equation}
 p_j\big(\sigma_j(a), \Delta(a)\big) = \frac{1}{\sigma_j(a)} \left(1 - e^{-j \, \sigma_j(a)\Delta(a)}
 \sum_{m=1}^{j} \frac{\big(j \, \sigma_j(a)\Delta(a)\big)^m}{m!} \right)
\end{equation}

Eq.$(34)$ can be written in a more feasible form for future calculations such as
\begin{equation}
 p_j\big(\sigma_j(a), \Delta(a)\big) = \frac{1}{\sigma_j(a)} \left(1-\frac{\Gamma\big(j+1, j\,\sigma_j(a)\Delta(a)\big)}{\Gamma(j+1)}
 \right)
\end{equation}
where
\begin{equation}
 \Gamma\big(j+1\big) = \int_0^{\infty} e^{-z} z^j \mathrm{d}z
\end{equation}
is the Gamma function \cite{abramowitz}. $\Gamma(j+1) = j!$ here, since $j$ is an integer. And
\begin{equation}
 \Gamma\big(j+1, j\,\sigma_j(a)\Delta(a)\big) = \int_0^{j\,\sigma_j(a)\Delta(a)} e^{-z} z^j \mathrm{d}z
\end{equation}
is related to the Incomplete Gamma function \cite{abramowitz}.

One should note that the real representations of probability distribution functions should be given by the restriction
$0 \leq p_j\big(\sigma(a),\Delta(a)\big) \leq 1$ on the surfaces $p_j\big(\sigma(a),\Delta(a)\big)$. In fig.2 we present these
surfaces for several $j$-values and the planes $p_j\big(\sigma(a),\Delta(a)\big) = 1$ and $p_j\big(\sigma(a),\Delta(a)\big) = 0.06$.
\begin{figure}[!hbt]
 \centering
 \includegraphics[width=0.66\linewidth]{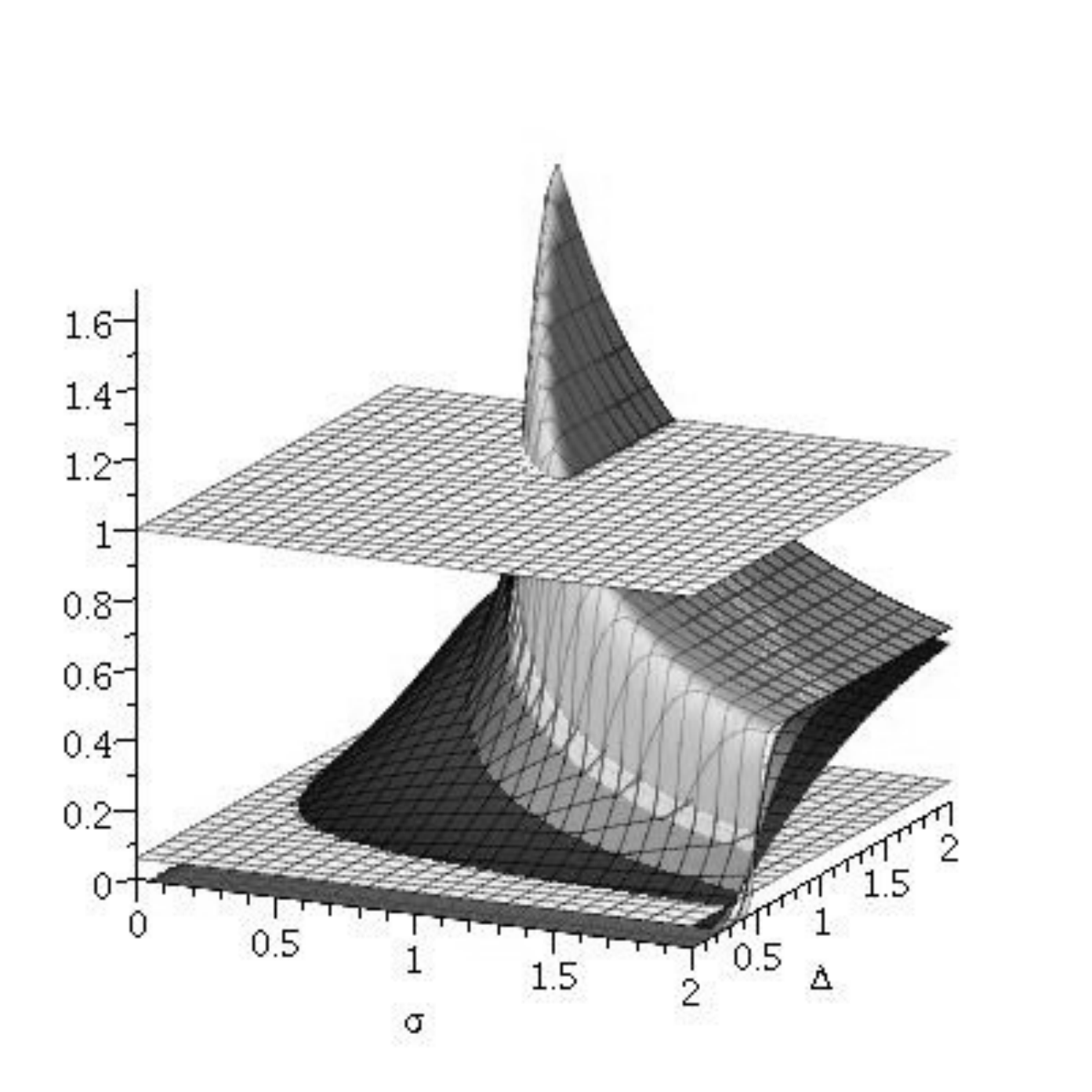}
 \caption{The surfaces $p_j\big(\sigma(a),\Delta(a)\big)$ for $j = 1, 6, 26, 27, 35, 200$ and the planes
 $p_j\big(\sigma(a),\Delta(a)\big) = 1$, $p_j\big(\sigma(a),\Delta(a)\big) = 0.06$.}
\end{figure}

\section{The Domain of the variables $\sigma(a)$, $\Delta(a)$. The Pattern Recognition based on Level Curves}
In the present section we set up the experimental scenario of the paper as well as the fundamental equations for the statistical
treatment of data. We will gave on table 1 below an example of probability distribution of amino acid occurrence for the $a =$ A
amino acid of the Pfam family PF01051 on a $(100\, \mathrm{x} \,200)$ block. The set $J$ of $j$-values $(j=1,2,\ldots,200)$ is
partitioned into subsets $J_s$ of values $j_s$. The $j_s$-values belonging to a subset $J_s$ do correspond to a unique constant value
$M_{J_s}\big(\sigma_{j_s}(A),\Delta(A)\big)$ of the function $p_{j_s}\big(\sigma_{j_s}(A),\Delta(A)\big)$, as
\begin{equation}
 p_{j_s}\big(\sigma_{j_s}(A),\Delta(A)\big) = M_{J_s}\big(\sigma_{j_s}(A),\Delta(A)\big) = \frac{n_{J_s}(A)}{100}
\end{equation}
where $n_{J_s}(A)$ is the number of occurrences of the $a =$ A amino acid on each subset $J_s$ of the $j_s$-values. These
values do correspond to the level curves of the surfaces $p_j\big(\sigma(a),\Delta(a)\big)$.

\begin{table}[!hbt]
 \begin{center}
  \caption{The $j_s$ values on each row of the second column belong to the subset $J_s$ of values for defining a level curve of the
  surfaces $p_{j_s}\big(\sigma(a),\Delta(a)\big)$. Data obtained from a $(100\, \mathrm{x} \,200)$ block as a representative of
  the Pfam family PF01051.}
  \begingroup
  \everymath{\scriptstyle}
  \tiny
   \begin{tabular}{|c|c|}
    \hline
    $M_{J_s}\big(\sigma_{J_s}(A),\Delta(A)\big)$ & $J_s$ \\
    \hline
    $0$ & $17, 18, 57, 179$ \\
    \hline
    $1/100$ & $2, 3, 12, 38, 79, 88, 97, 111, 120, 157, 166, 178, 180$ \\
    \hline
    $2/100$ & $5, 36, 82, 92, 125, 148, 172, 173, 175, 176$ \\
    \hline
    $3/100$ & $7, 41, 45, 46, 48, 58, 81, 100, 105, 119, 124, 133, 136, 147, 150, 151, 155, 161, 162, 165, 171, 187, 197, 199$ \\
    \hline
    \multicolumn{1}{|c}{\multirow{2}{*}{$4/100$}} & \multicolumn{1}{|c|}{$4, 8, 23, 24, 25, 60, 61, 73, 74, 75, 78, 80, 83, 84, 86,
    89, 90, 91, 99, 107, 112, 113, 123, 126, 127, 129, 131,$}\\
    \multicolumn{1}{|c}{} & \multicolumn{1}{|c|}{$135, 142, 145, 149, 154, 156, 168, 170, 184, 188, 192, 193, 198$}\\
    \hline
    $5/100$ & $9, 11, 19, 34, 40, 44, 56, 62, 94, 96, 101, 102, 104, 106, 121, 134, 140, 158, 159, 181, 186, 189$\\
    \hline
    $6/100$ & $1, 6, 26, 27, 35, 37, 39, 51, 54, 95, 109, 115, 118, 122, 138, 146, 164, 174, 177, 194, 200$\\
    \hline
    $7/100$ & $13, 20, 52, 55, 63, 65, 68, 70, 76, 87, 103, 130, 152, 167, 169, 183, 185, 191$\\
    \hline
    $8/100$ & $22, 28, 47, 85, 93, 110, 114, 117, 139, 144, 153, 160, 163, 182, 190, 196$\\
    \hline
    $9/100$ & $15, 32, 42, 43, 49, 59, 66, 72, 98, 108, 137$\\
    \hline
    $10/100$ & $116, 143$\\
    \hline
    $11/100$ & $64, 67, 141$\\
    \hline
    $12/100$ & $21, 30, 195$\\
    \hline
    $13/100$ & $50, 128, 132$\\
    \hline
    $14/100$ & $33$\\
    \hline
    $15/100$ & $31, 71$\\
    \hline
    $16/100$ & $77$\\
    \hline
    $17/100$ & $53$\\
    \hline
    $18/100$ & $10, 69$\\
    \hline
    $21/100$ & $16$\\
    \hline
    $26/100$ & $14$\\
    \hline
    $30/100$ & $29$\\
    \hline
   \end{tabular}
   \endgroup
  \end{center}
\end{table}

In figure 3, we represent the histogram corresponding to the probabilities of occurrence of the $a =$ A amino acid which have been
obtained from the family PF01051. This should be compared to figure 4 in which we advance the representation of the level curves of
this amino acid distribution. It seems to exist a clear advantage of the last representation over the histogram and we will intend to
represent in the same form the distribution of other amino acids from selected families.
\begin{figure}[!htb]
 \centering
 \includegraphics[width=0.375\linewidth]{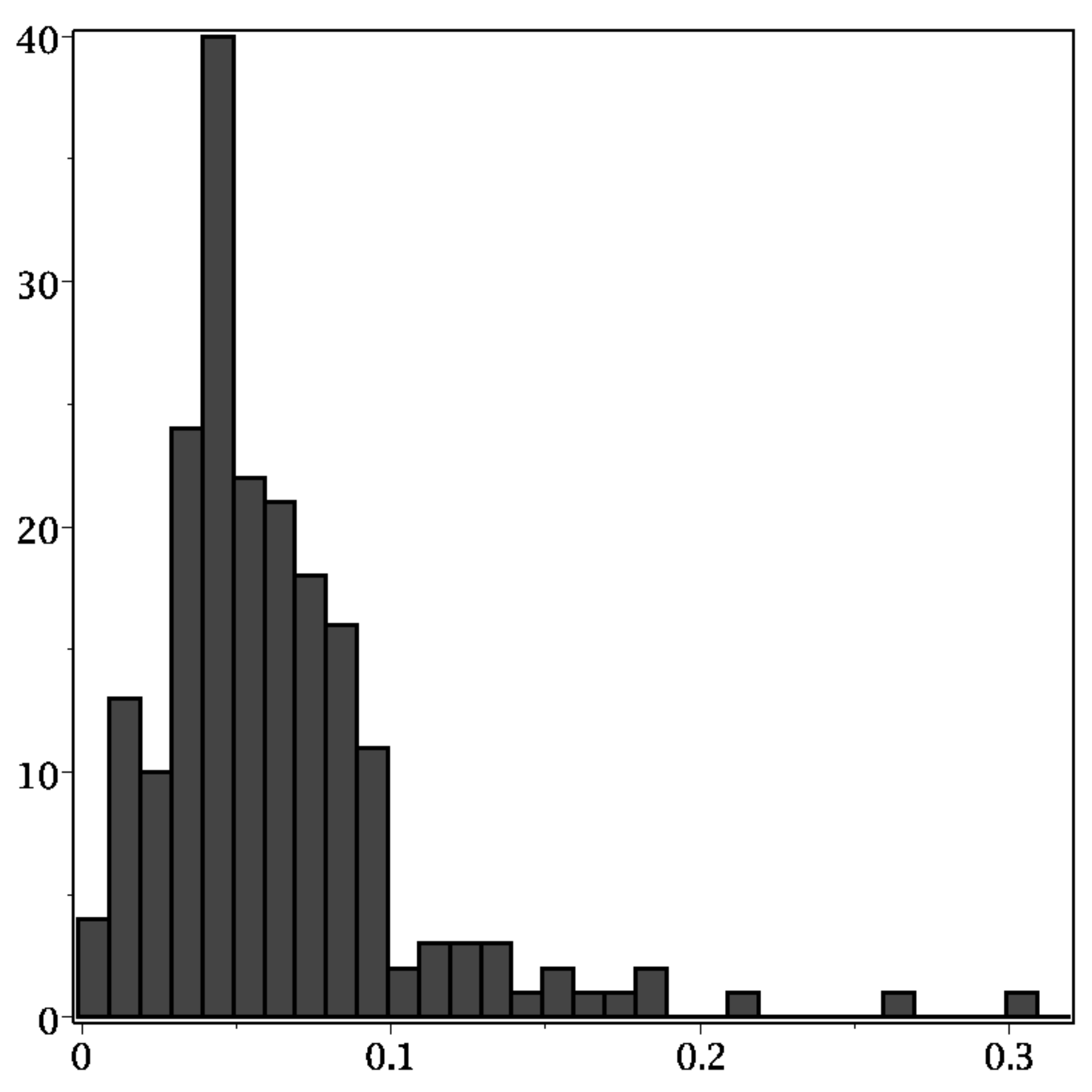}
 \caption{The histogram of the distribution corresponding to Table 1. Amino acid A, Pfam family PF01051.}
\end{figure}

We now undertake a detailed analysis which led to the pictorial representation of fig.4 in order to determine the most convenient
domain for the variables $\sigma(a)$, $\Delta(a)$.
\begin{figure}[!hbt]
 \centering
 \includegraphics[width=0.375\linewidth]{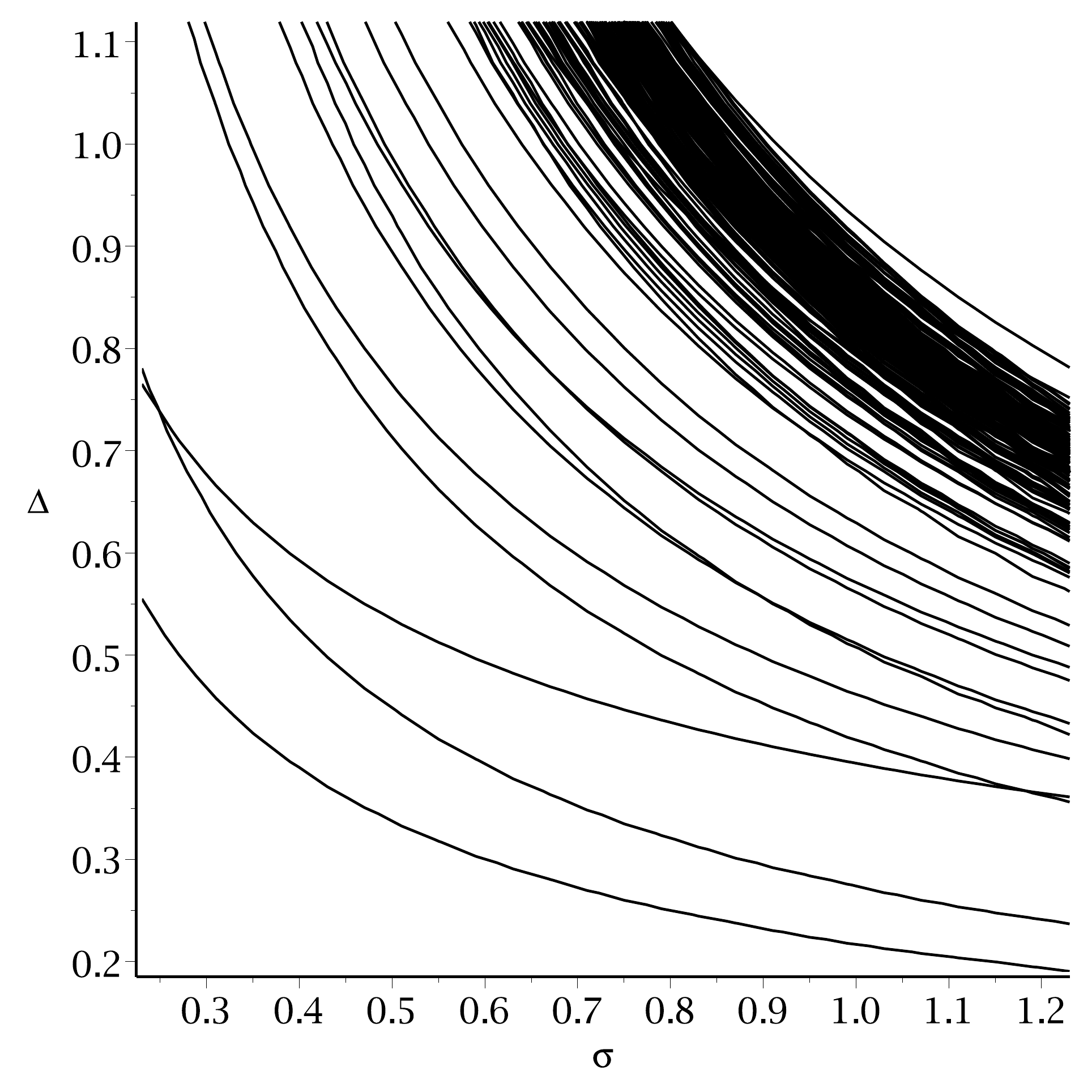}
 \caption{The level curves corresponding to the distribution given in Table 1 and Fig 3. Amino acid A, Pfam family PF01051.}
\end{figure}

First of all, we note that the usual analysis of extrema of the surfaces given by eq.$(35)$ is inconclusive, since from the
equations:
\begin{equation}
 \frac{\partial p_j\big(\sigma_j(a), \Delta(a)\big)}{\partial \sigma_j(a)} = 0 \,\, ; \quad
 \frac{\partial p_j\big(\sigma_j(a), \Delta(a)\big)}{\partial \Delta(a)} = 0
\end{equation}
we get as the only one solution the point $\big(\sigma(a), \Delta(a)\big) = (0,0)$ and we have:
\begin{equation}
 \mathrm{Hess} (0,0) = \det
 \begin{pmatrix}
  \frac{\partial^2 p_j\big(\sigma_j(a), \Delta(a)\big)}{\partial^2 \sigma_j(a)} \bigg|_{(0,0)} &
  \frac{\partial^2 p_j\big(\sigma_j(a), \Delta(a)\big)}{\partial \sigma_j(a) \partial \Delta(a)} \bigg|_{(0,0)} \\ \relax
  \frac{\partial^2 p_j\big(\sigma_j(a), \Delta(a)\big)}{\partial \sigma_j(a) \partial \Delta(a)} \bigg|_{(0,0)} &
  \frac{\partial^2 p_j\big(\sigma_j(a), \Delta(a)\big)}{\partial^2 \Delta(a)} \bigg|_{(0,0)}
 \end{pmatrix}
 = 0
\end{equation}

We then proceed to an alternative analysis in order to characterize the 2-dimensional domain of variables $\sigma(a)$, $\Delta(a
)$ for the modelling process. This is based on the study of the intersection of the $p_j\big(\sigma(a), \Delta(a)\big)$ surfaces
by horizontal and vertical planes.

The horizontal planes $h_j(a)$ will determine the level curves which are given by:
\begin{equation}
 \resizebox{.9\hsize}{!}{$p_j\big(\sigma_j(a), \Delta(a)\big) = M_j(a) = \frac{1}{\sigma_j(a)}\left(1- 
 \frac{\Gamma\big(j+1, j\,\sigma_j(a)\Delta(a)\big)}{\Gamma(j+1)}\right) \, \forall a \,,\, 1 \leq j \leq n$}
\end{equation}
where $M_j(a)$ corresponds to a characteristic constant value associated to the subset of $j$-values of the specific
amino acid according to the example given in Table 1.

From eq.$(41)$, we can write:
\begin{equation}
 \mathrm{d}\sigma_j(a)\frac{\partial p_j\big(\sigma_j(a), \Delta(a)\big)}{\partial \sigma_j(a)} +
 \mathrm{d}\Delta(a)\frac{\partial p_j\big(\sigma_j(a), \Delta(a)\big)}{\partial \Delta(a)} = 0
\end{equation}

The local minimum points $\big(\sigma_{j\,\min}(a),\Delta_{\min}(a)\big)$ of the level curves are obtained from eq.$(41)$ and from
\begin{equation}
 \frac{\mathrm{d}\Delta(a)}{\mathrm{d}\sigma_j(a)} = 0 = \sigma_j(a)\left(\Gamma(j+1) - \big(j\,\Delta(a)\big)^{j+1}
 \big(\sigma_j(a)\big)^j e^{-j\,\sigma_j(a)\Delta(a)}\right)
\end{equation}
We now consider the vertical planes $V_j(a)$, associated to a generic value $\delta_j(a)$. On each vertical plane, there is a
vertical curve such that its local maximum is given by $\big(\sigma_{j\,\max}(a),\delta_{j\,\max}(a)\big)$.

These vertical curves are given by
\begin{equation}
 p_j\big(\sigma_j(a), \delta_j(a)\big) = \frac{1}{\sigma_j(a)}\left(1- 
 \frac{\Gamma\big(j+1, j\,\sigma_j(a)\delta_j(a)\big)}{\Gamma(j+1)}\right)
\end{equation}
and their local maxima can be obtained from eq.$(44)$, and from
\begin{equation}
 \resizebox{.92\hsize}{!}{$\frac{\partial p_j\big(\sigma_j(a), \Delta(a)\big)}{\partial \sigma_j(a)} = 0 = -\frac{1}{\sigma_j^2
 (a)} \left(1 - \frac{\Gamma\big(j+1, j\,\sigma_j(a)\Delta_j(a)\big) + \big(j\,\sigma_j(a)\delta_j(a)\big)^{j+1} e^{-j\,\sigma_j
 (a)\delta_j(a)}}{\Gamma(j+1)}\right)$}
\end{equation}

From eqs.$(41)$, $(43)$-$(45)$, we can write:
\begin{gather}
 M_j(a) = \frac{1}{\sigma_{j_{\min}}(a)}\left(1- \frac{\Gamma\big(j+1, j\,\sigma_{j_{\min}}(a)\Delta_{\min}(a)\big)}{\Gamma(j+1)}
 \right)\\
 \Gamma(j+1) - \big(j\,\Delta_{\min}(a)\big)^{j+1}\big(\sigma_{j_{\min}}(a)\big)^j e^{-j\,\sigma_{j_{\min}}(a)\Delta_{\min}(a)}=0\\
 p_j(a) = \frac{1}{\sigma_{j_{\max}}(a)}\left(1- \frac{\Gamma\big(j+1, j\,\sigma_{j_{\max}}(a)\delta_{j_{\max}}(a)\big)}{\Gamma
 (j+1)}\right)\\
 \Gamma(j+1)p_j(a) - \big(j\,\delta_{j_{\max}}(a)\big)^{j+1}\big(\sigma_{j_{\max}}(a)\big)^j e^{-j\,\sigma_{j_{\max}}(a)\Delta
 _{\max}(a)}=0
\end{gather}

Some remarks should be done before we proceed to determine the points\\ $\big(\sigma_{j_{\min}}(a)\Delta_{\min}(a)\big)$,
$\big(\sigma_{j_{\max}}(a)\delta_{j_{\max}}(a)\big)$ from eqs.$(46)$-$(49)$. From eq.$(45)$ we see that the product
$\sigma(a)\Delta(a)$ does not depend on the amino acid \textbf{a}, or
\begin{equation}
 \sigma_j(a)\delta_j(a) = f(j) \,, \quad \forall (a)
\end{equation}

Let us suppose the ordering of $\Delta_j(a)$ values for different amino acids but at the same $j$-th column of the $(m\,
\mathrm{x}\,n)$ block,
\begin{equation}
 \delta_j(a) \geq \delta_j(b) \geq \delta_j(c)
\end{equation}
We then have from eq.$(50)$
\begin{equation}
 \sigma_j(a) \leq \sigma_j(b) \leq \sigma_j(c)
\end{equation}

An example of level curves corresponding to these inequalities can be seen at fig.5.

Let us also suppose that there is an amino acid $\mathbf{a'}$ such that for two columns $j_1$, $j_2$ we have:
\begin{equation}
 \delta_{j_1}(a') \geq \delta_j(a) \geq \delta_{j_2}(a') \,, \quad j_1 \leq j \leq j_2
\end{equation}
From eq.$(50)$, we can also write:
\begin{equation}
 \sigma_{j_1}(a') \leq \sigma_j(a) \leq \sigma_{j_2}(a') \,, \quad j_1 \leq j \leq j_2
\end{equation}

\begin{figure}[!htb]
 \centering
 \includegraphics[width=0.5\linewidth]{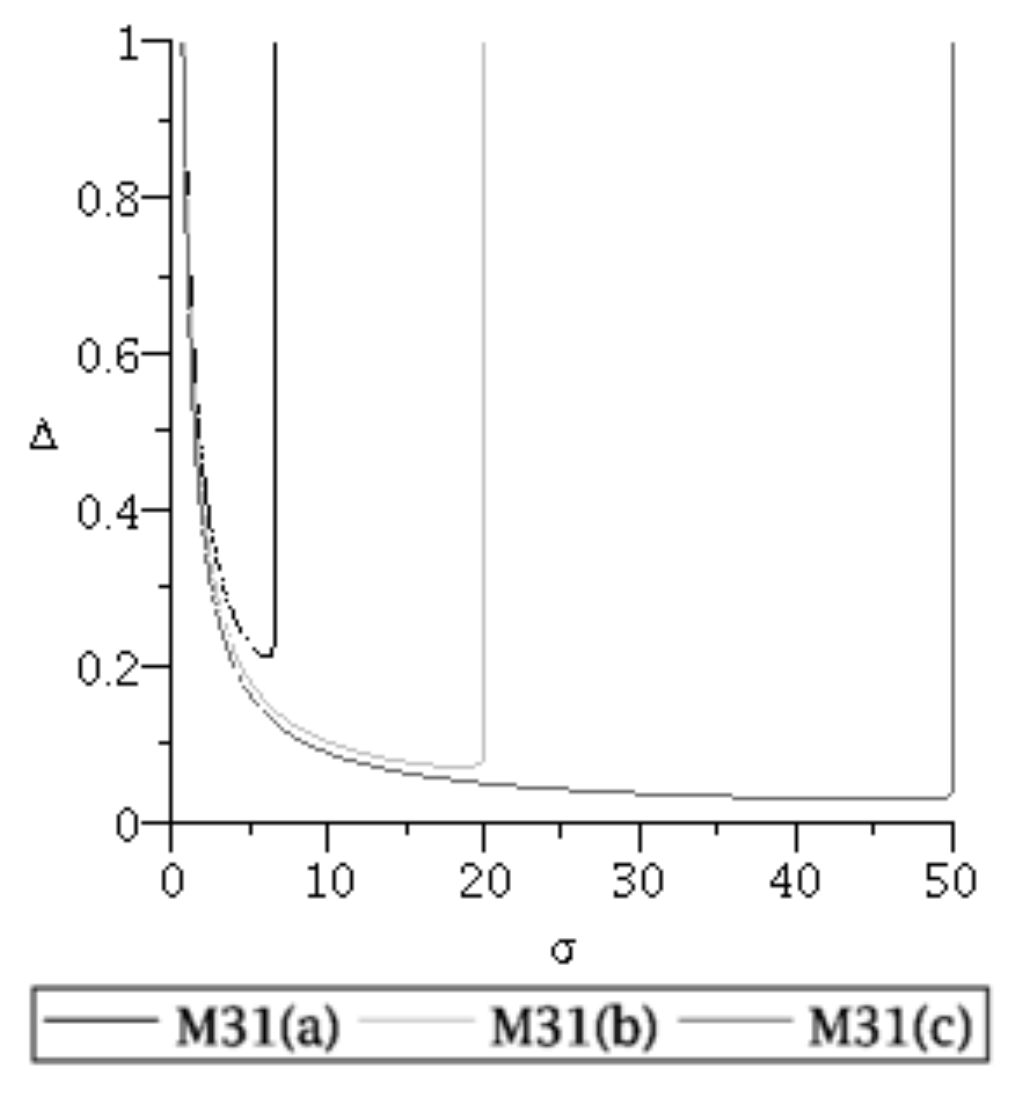}
 \caption{Three level curves corresponding to the $j = 31$ value and the amino acids $a = A$, $b = C$, $c = D$ of the 
 $100\,\mathrm{x}\,200$ block of the PF01051 family.}
\end{figure}

We can take $j_2 = j_{\max} = 200$ and $j_1 = j_{\min} = 1$ in the case of a $100\,\mathrm{x}\,200$ block. In fig.6 an example
of the level curves for $j = 55$ and the amino acids $a' = K$, $a = L$ from PF01051 family.

\begin{figure}[hbt]
 \centering
 \includegraphics[width=0.5\linewidth]{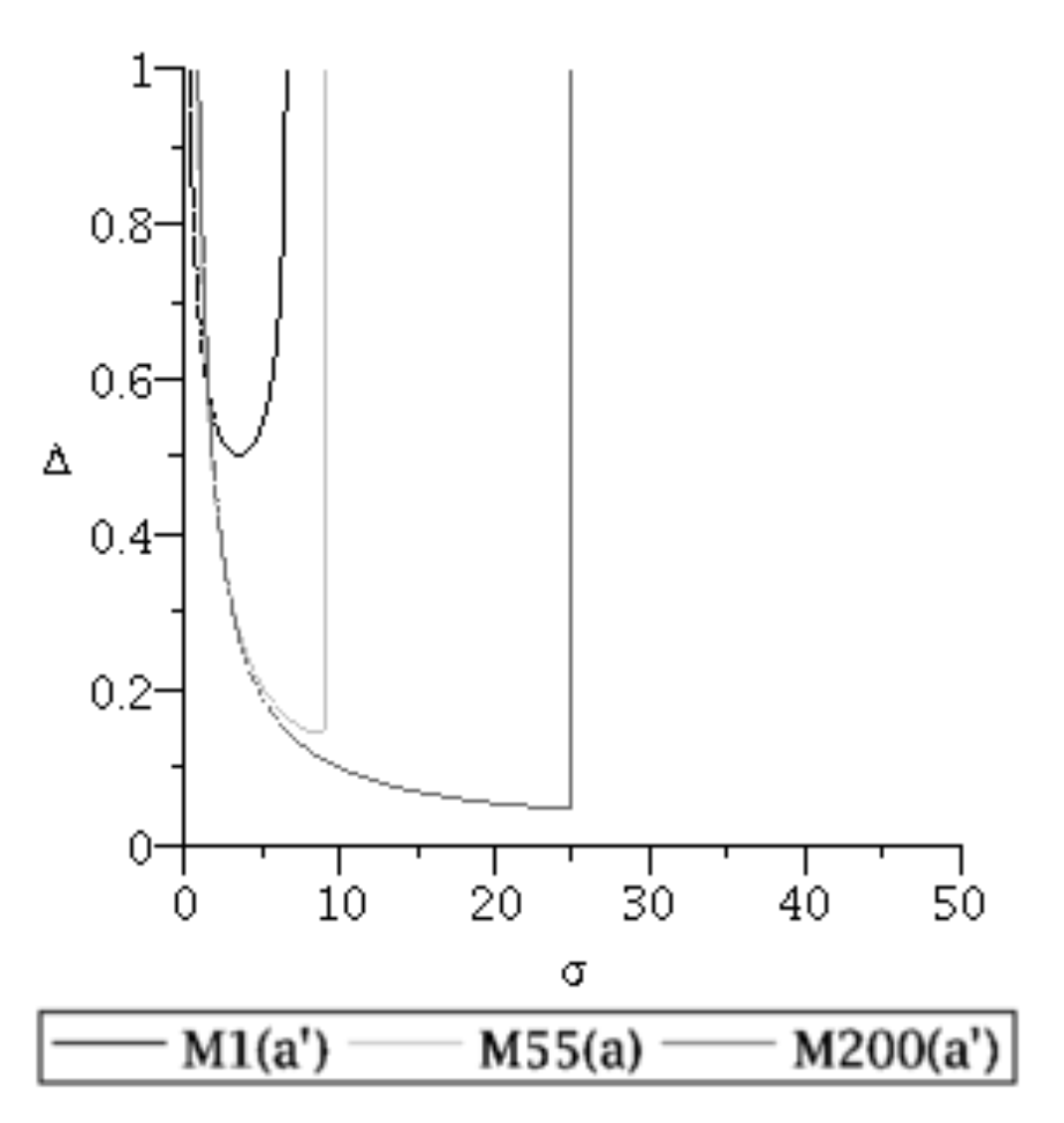}
 \caption{Three level curves corresponding to the $j_1 = 1$, $j_2 = 200$, $j = 55$, $a' = K$, $a = L$.}
\end{figure}

We can then look for the two-dimensional domain of the process leading to pattern recognition. We look for $j_1$, $j_2$, $j_3$
such that
\begin{equation}
 \sigma_{j_1\,\min}(a) \geq \sigma(a) \geq \sigma_{j_2\,\max}(a)
\end{equation}
and
\begin{equation}
 \delta_{j_3\,\max}(a) \geq \Delta(a) \geq \Delta_{\min}(a)
\end{equation}

We are now ready to proceed to the determination of the values $\sigma_{j_{\min}}(a)$, $\Delta_{\min}(a)$, $\sigma_{j_{\max}}(a)$,
$\delta_{j_{\max}}(a)$ from eqs.$(46)$-$(49)$.

First of all we note that eq.$(47)$ can be also written as:
\begin{equation}
 \resizebox{.92\hsize}{!}{$-\frac{j}{j+1}\sigma_{j_{\min}}(a)\Delta_{\min}(a) e^{\frac{-j\,\sigma_{j_{\min}}(a)\Delta_{\min}}{j+1}}
 = -\frac{\sigma_{j_{\min}}(a)}{j+1}\left(\Gamma(j+1)\sigma_{j_{\min}}^{-j}(a)\right)^{\frac{1}{j+1}}$}
\end{equation}
and this should be compared to
\begin{equation}
 W(z)e^{W(z)} = z
\end{equation}
where $W(z)$ is the Lambert W Non-injective function [13]. We then have:
\begin{equation}
 \Delta_{\min}(a) = -\frac{(j+1)}{j\,\sigma_{j_{\min}}(a)}W\left(-\frac{\sigma_{j_{\min}}(a)}{j+1}\big(\Gamma(j+1)
 \sigma_{j_{\min}}^{-j}(a)\big)^{\frac{1}{j+1}}\right)
\end{equation}

Analogously, we can write from eq.$(49)$:
\begin{equation}
 \delta_{j\,\max}(a) = -\frac{(j+1)}{j\,\sigma_{j\,\max}(a)}W\left(-\frac{\sigma_{j\,\max}(a)}{j+1}\big(\Gamma(j+1)p_j(a)
 \sigma_{j\,\max}^{-j}(a)\big)^{\frac{1}{j+1}}\right)
\end{equation}

The self-consistency of the system of equations $(46)$, $(59)$, $(48)$, $(60)$, will be proved in the following way:\\
If we assume that $M_j(a)$ is given, we will get $\sigma_{j\,\min}(a)$ from eqs.$(46)$ and $(59)$. Eq.$(59)$ will then give
$\Delta_{\min}(a)$. If we assume that $p_j(a)$ is given we will get $\sigma_{j\,\max}(a)$ from eqs.$(48)$ and $(60)$. We will
then get $\delta_{j\,\max}(a)$ from eq.$(60)$.

A special case is able to motivate the understanding of the developments above. We consider $\delta_{j\,\max}(a) = \Delta_{\min}
(a)$ and $\sigma_{j\,\max}(a) = \sigma_{j\,\min}(a)$. This can be seen at fig.7 below. The corresponding vertical and horizontal
curves will meet at point P.
\begin{figure}[!htb]
 \centering
 \includegraphics[width=0.75\linewidth]{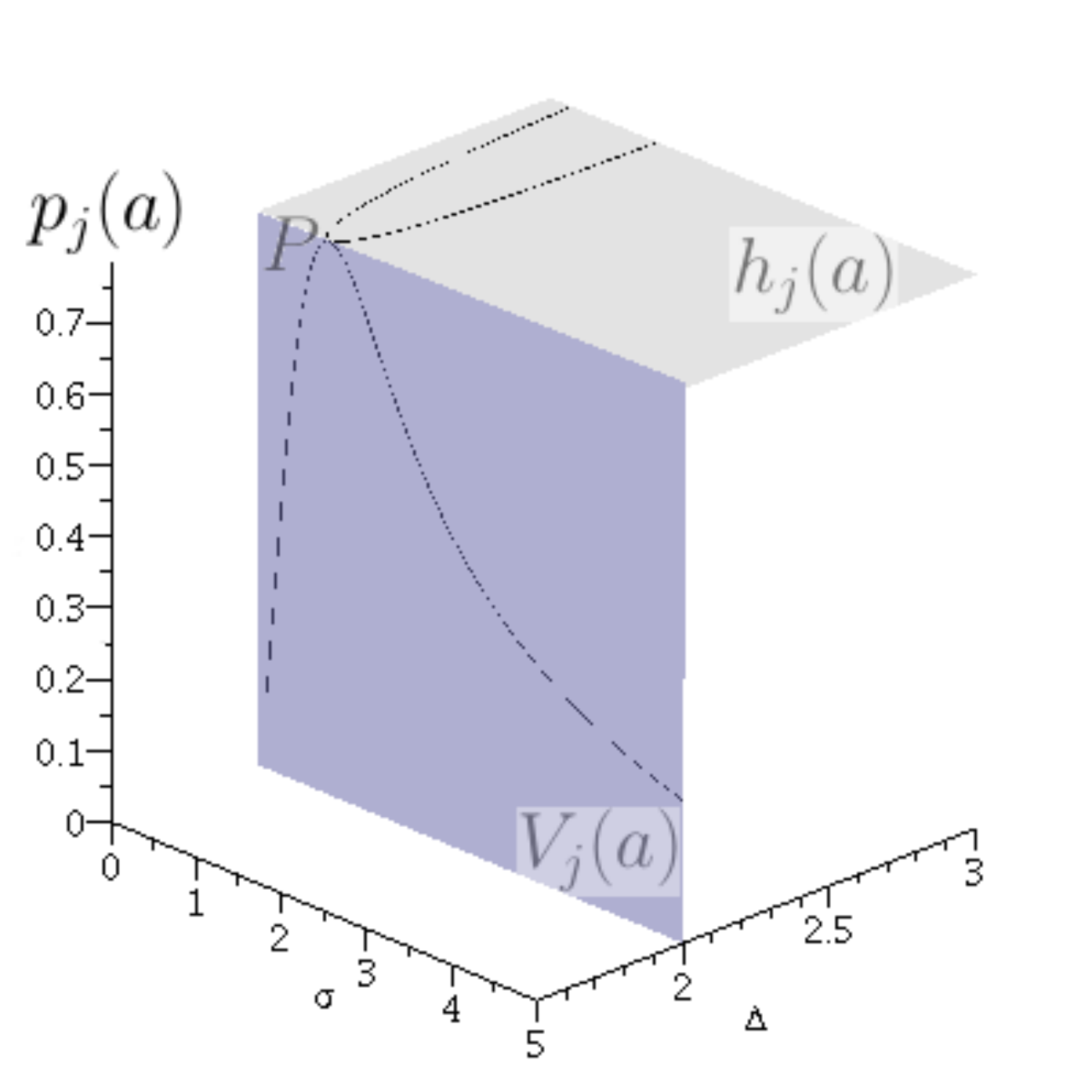}
 \caption{Sketch of the level and vertical curves corresponding to the case $\sigma_{j\,\max}(a) = \sigma_{j\,\min}(a)$ and 
 $\delta_{j\,\max}(a) = \Delta_{\min}(a)$ for a generic amino acid $\mathrm{a}$ and a $j$-value $1 \leq j \leq n$. $h_j(a)$ and
 $V_j(a)$ are horizontal and vertical planes, respectively.}
\end{figure}

The intersection of the surface $p_j\big(\sigma_j(a), \Delta(a)\big)$, eq.$(46)$ with the plane $M_n = 1$, will lead to
\begin{equation}
 1 - \sigma_{n\,\min}(a) = \frac{\Gamma\big(n+1, n\,\sigma_{n\,\min}(a)\Delta_{\min}(a)\big)}{\Gamma(j+1)}
\end{equation}

We can write from eqs.$(61)$ and $(34)$, if $n \gg 1$
\begin{equation}
 1 \approx \frac{1}{\sigma_{n\,\min}(a)}
\end{equation}
By applying the Stirling approximation as given by
\begin{equation}
 n! \approx e^{-n}(n)^n \,, \quad n \gg 1
\end{equation}
We get from eqs.$(47)$ and $(62)$:
\begin{equation}
 e^{-n}(n)^n - e^{-n\,\Delta_{\min}(a)}\big(n\,\Delta_{\min}(a)\big)^{n+1} \approx 0
\end{equation}
For $n = 200$, we can write from eqs.$(62)$, $(64)$:
\begin{equation}
 \sigma_{n\,\min}(a) \approx 1.0000 \, ; \quad \Delta_{\min}(a) \approx 1.2538
\end{equation}

The intersection of the surface $p_j\big(\sigma_j(a),\delta_j(a)\big)$, eq.$(48)$, with the plane $M_1 = 1$ will lead to
\begin{equation}
 1 - \sigma_{j\,\max(a)} = \Gamma\big(2, \sigma_{1\,\max}(a)\delta_{1\,\max}\big)
\end{equation}
From eq.$(34)$, the last equation can be also written:
\begin{equation}
 -\big(1 + \sigma_{1\,\max}(a)\delta_{1\,\max}(a)\big) e^{-\big(1 + \sigma_{1\,\max}(a)\delta_{1\,\max}(a)\big)} =
 -\big(1 - \sigma_{1\,\max}(a)\big) e^{-1}
\end{equation}
After comparison of eq.$(67)$ with eq.$(58)$, we write,
\begin{equation}
 \delta_{1\,\max}(a) = -\frac{1}{\sigma_{1\,\max}}\bigg(1+ W\Big(\big(\sigma_{1\,\max}(a)-1\big)e^{-1}\Big)\bigg)
\end{equation}
We also have from eq.$(60)$,
\begin{equation}
 \delta_{1\,\max}(a) = -\frac{2}{\sigma_{1\,\max}} W\left(\pm \frac{\big(\sigma_{1\,\max}(a)\big)^{\frac{1}{2}}}{2}\right)
\end{equation}
We then get from eqs.$(68)$, $(67)$:
\begin{equation}
 \sigma_{1\,\max}(a) = 0.5352 \, ; \quad \delta_{1\,\max}(a) = 3.3509
\end{equation}

By using the results of eqs.$(65)$, $(70)$ into eqs.$(55)$, $(56)$, we should have $j_1 = n$, $j_2 = 1$, $j_3 = 1$ and we write
for the domain of the variables $\sigma(a)$, $\Delta(a)$
\begin{equation}
 1.0000 \geq \sigma(a) \geq 0.5352 \, ; \quad 3.3509 \geq \Delta(a) \geq 1.2538
\end{equation}

\section{Final Numerical Calculations and Graphical Representation}

In this section we show the proposed graphical representation of the probability occurrences of all amino acids from some
selected protein families. These representations do correspond to the level curves of the surface of probability distribution,
eq.$(39)$ and an example has been already presented on fig.4 of a cartesian representation for the $a =$ A amino acid of family
PF01051. All the necessary essential steps have been derived in the previous sections and we stress once more that the cartesian
representations of the level curves is based on the partition of the $j$-values $(1\,\leq j \leq\,n)$ into subsets $J_s$. Each
$j$-values belonging to a $J_s$ subset is then associated to a constant value $M_{j_s}\big(\sigma(a), \Delta(a)\big)$ of the
probability distribution function, according to eq.$(38)$.

In order to emphasize the usefulness and efficiency of the cartesian representation of level curves, we shall make a comparison of
probability occurrences in terms of histograms, figs.8, 9, 10 and level curves, figs.11, 12, 13 for the families PF01051, PF03399,
which belong to the same Clan CL00123 and PF00060 belonging to the Clan CL00030. We have chosen for representation of four rows and
five columns corresponding to table 2 below. The one-letter code for amino acids are ordered by decreasing
hydrophobicities \cite{sereda}, with cells to be read from top left to bottom right, at pH2.
\begin{table}[!hbt]
 \begin{center}
  \caption{The representation scheme of amino acids ordered according to decreasing hydrophobicities, top left to bottom right, at
  pH2.}
   \begin{tabular}{|c|c|c|c|c|}
    \hline
    $A_{00} = L$ & $A_{01} = I$ & $A_{02} = F$ & $A_{03} = W$ & $A_{04} = V$\\
    \hline
    $A_{10} = M$ & $A_{11} = C$ & $A_{12} = Y$ & $A_{13} = A$ & $A_{14} = T$\\
    \hline
    $A_{20} = E$ & $A_{21} = G$ & $A_{22} = S$ & $A_{23} = Q$ & $A_{24} = D$\\
    \hline
    $A_{30} = R$ & $A_{31} = K$ & $A_{23} = N$ & $A_{33} = H$ & $A_{34} = P$\\
    \hline
   \end{tabular}
  \end{center}
\end{table}

In order to introduce the cartesian representation of the level curves, we restrict the limits of the variables $\sigma(a)$,
$\Delta(a)$ of eq.$(72)$ to each cell of the array given in table 2. A generic element $A_{kl}$, $k = 0,1,2,3$, $l = 0,1,2,3,4$
is associated to the limits:
\begin{gather}
 \resizebox{.92\hsize}{!}{$\varphi \frac{\sigma_{1\,\max}(a)}{\sigma_{n\,\min}(a)-\sigma_{1\,\max}(a)} + l \leq
 \frac{\sigma(a)}{\sigma_{n\,\min}(a)-\sigma_{1\,\max}(a)} \leq \varphi \frac{\sigma_{1\,\max}(a)}{\sigma_{n\,\min}(a)
 -\sigma_{1\,\max}(a)} + (l+1)$}\\
 \resizebox{.92\hsize}{!}{$\varphi \frac{\delta_{1\,\max}(a)}{\delta_{1\,\max}(a)-\Delta_{\min}(a)} + (3-k) \leq
 \frac{\Delta(a)}{\delta_{1\,\max}(a)-\Delta_{\min}(a)} \leq \varphi \frac{\delta_{1\,\max}(a)}{\delta_{1\,\max}(a)-
 \Delta_{\min}(a)} + (4-k)$}
\end{gather}
where $\sigma_{1\,\max}(a)$, $\sigma_{n\,\min}(a)$, $\delta_{1\,\max}(a)$, $\Delta_{\min}(a)$ are given into eqs.$(65)$
and $(71)$ and $\varphi$ is an arbitrary non-dimensional parameter to be chosen as $\varphi = 0.2$ for obtaining a convenient
cartesian representation.

\newpage

\begin{figure}[H]
 \centering
 \includegraphics[width=0.85\linewidth]{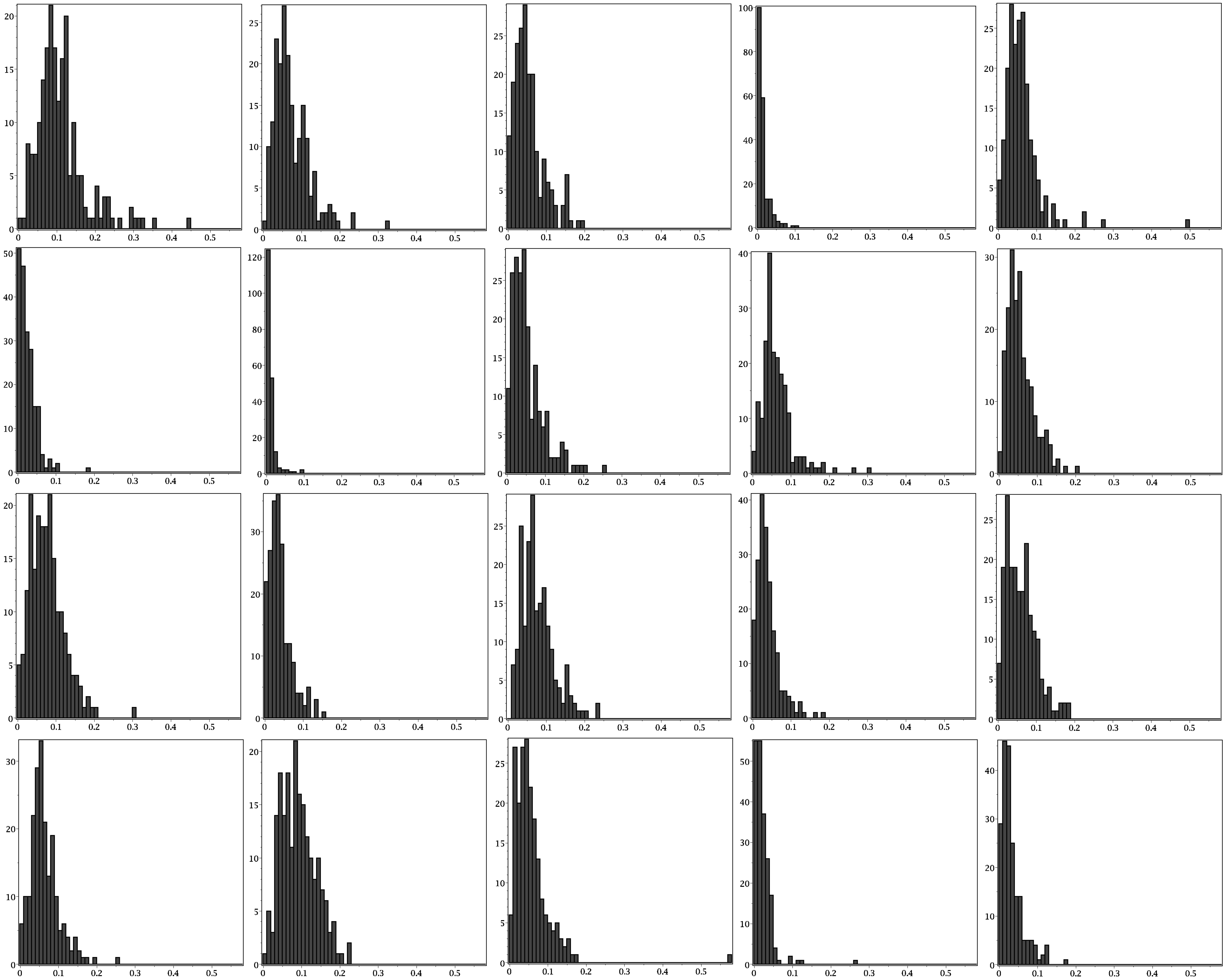}
 \caption{Histograms of the probability occurrences of amino acids for the Pfam protein family PF01051 from Clan CL00123.}
\end{figure}

\begin{figure}[H]
 \centering
 \includegraphics[width=0.85\linewidth]{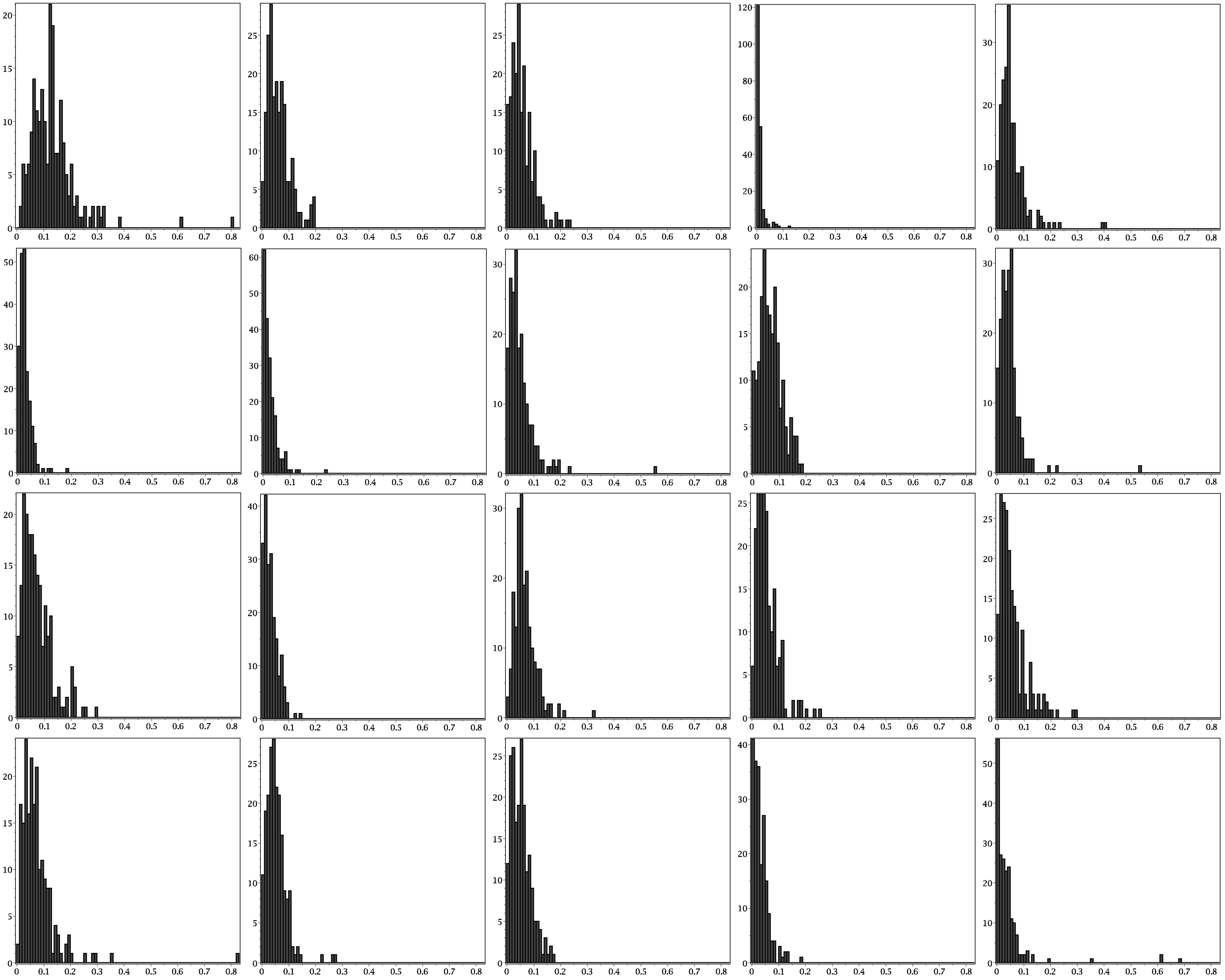}
 \caption{Histograms of the probability occurrences of amino acids for the Pfam protein family PF03399 from Clan CL00123.}
\end{figure}

\begin{figure}[H]
 \centering
 \includegraphics[width=0.85\linewidth]{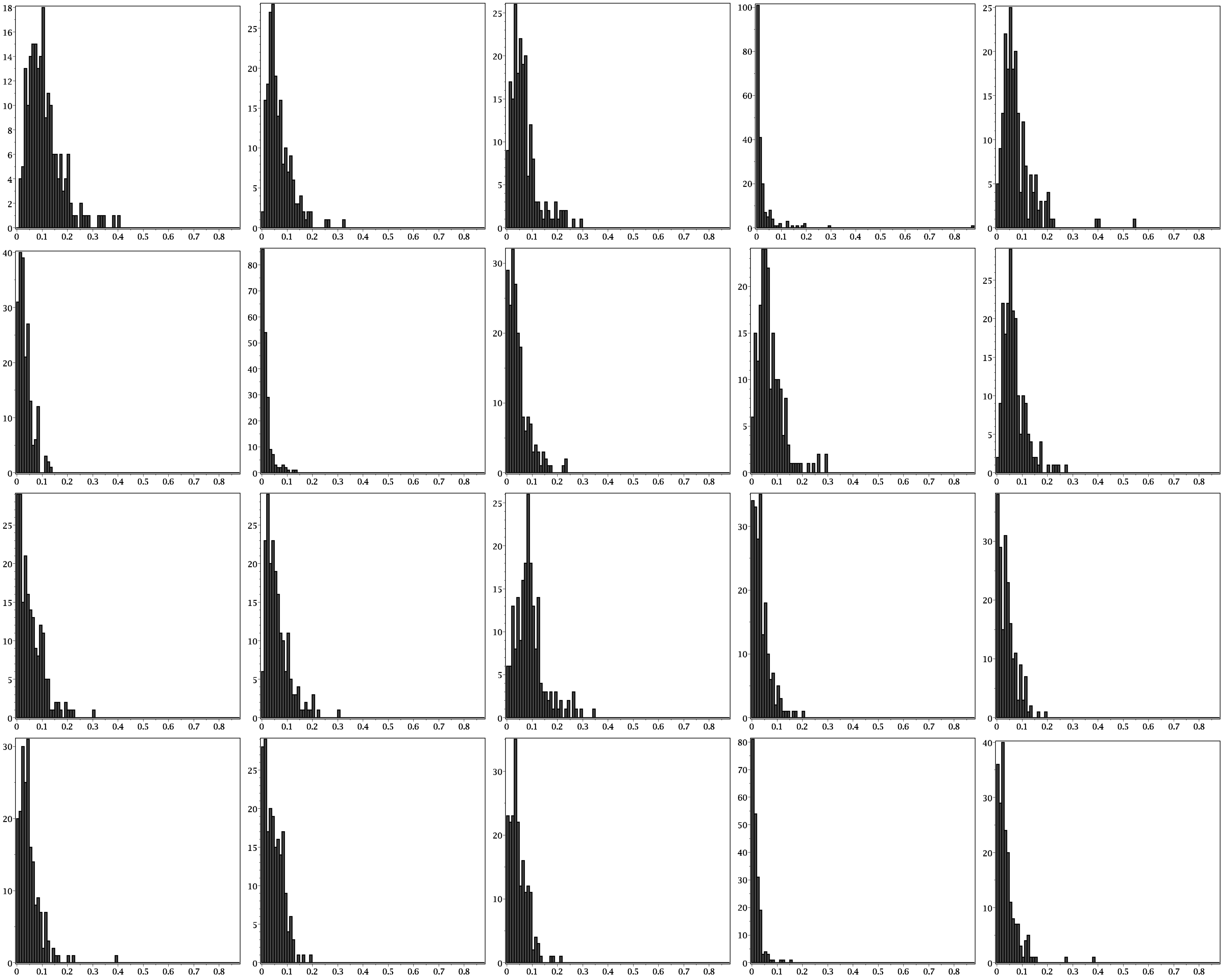}
 \caption{Histograms of the probability occurrences of amino acids for the Pfam protein family PF00060 from Clan CL00030.}
\end{figure}

\begin{figure}[H]
 \centering
 \includegraphics[width=0.85\linewidth]{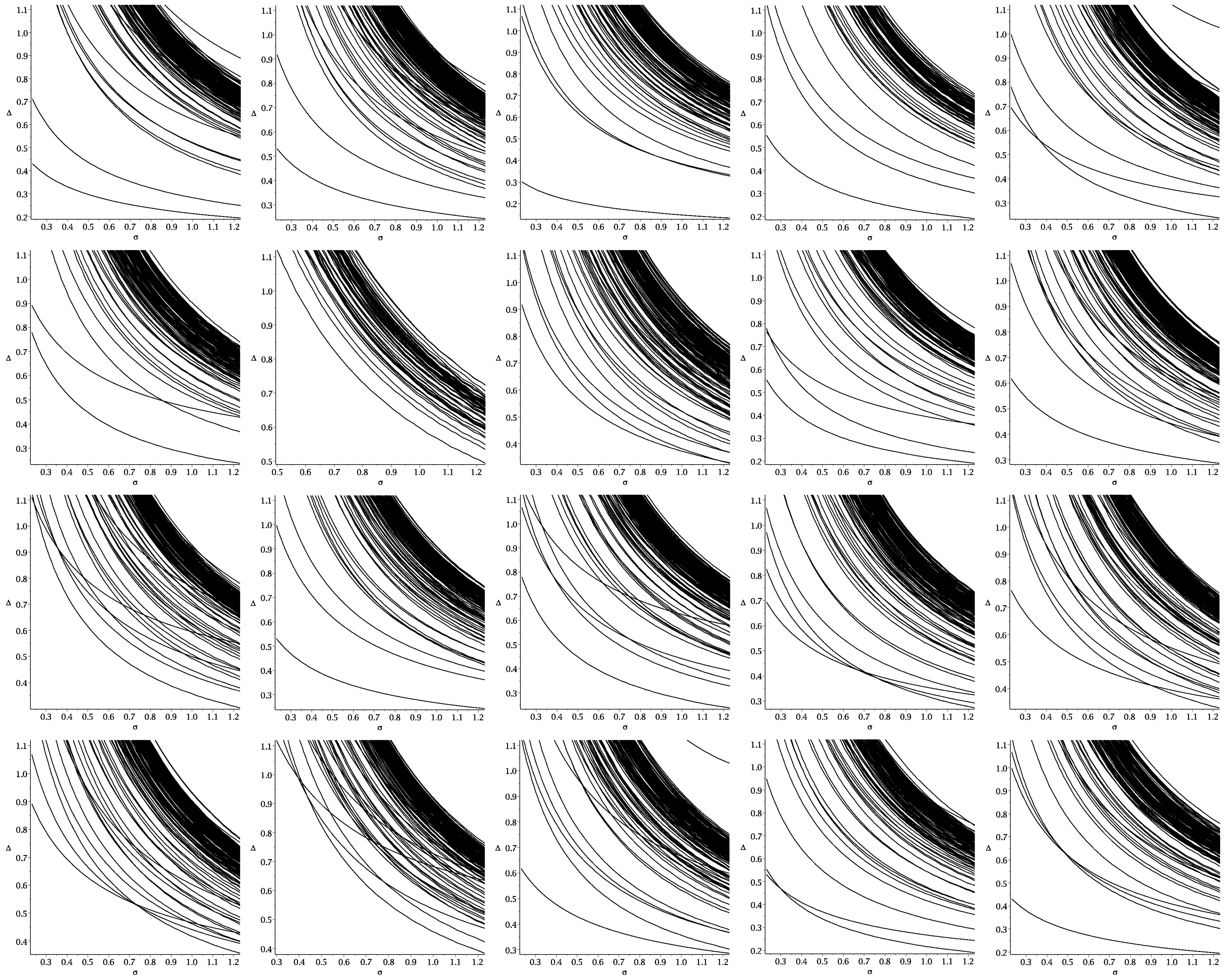}
 \caption{The level curves of the probability occurrences of amino acids for the Pfam protein family PF01051 from Clan CL00123.}
\end{figure}

\begin{figure}[H]
 \centering
 \includegraphics[width=0.85\linewidth]{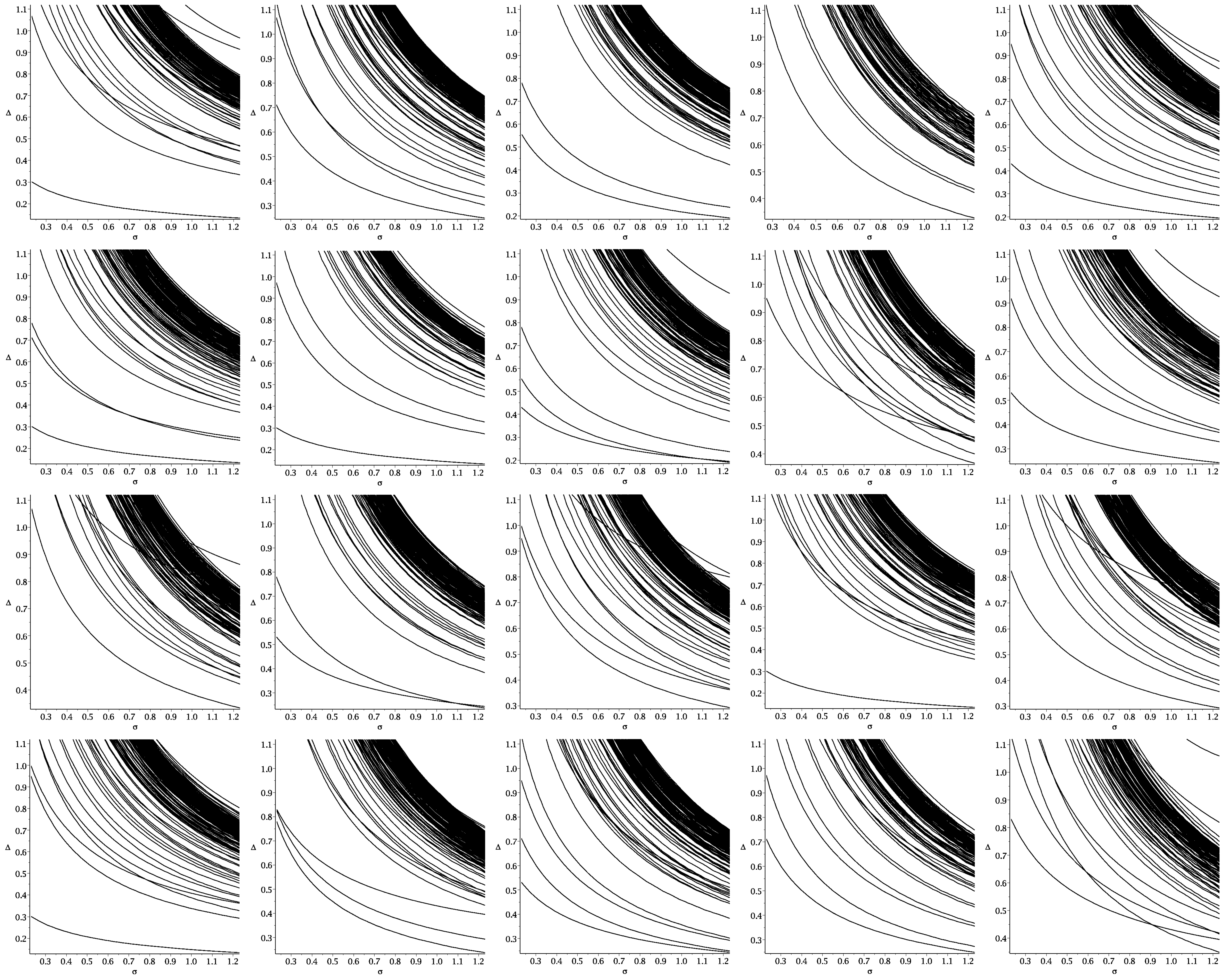}
 \caption{The level curves of the probability occurrences of amino acids for the Pfam protein family PF03399 from Clan CL00123.}
\end{figure}

\begin{figure}[H]
 \centering
 \includegraphics[width=0.85\linewidth]{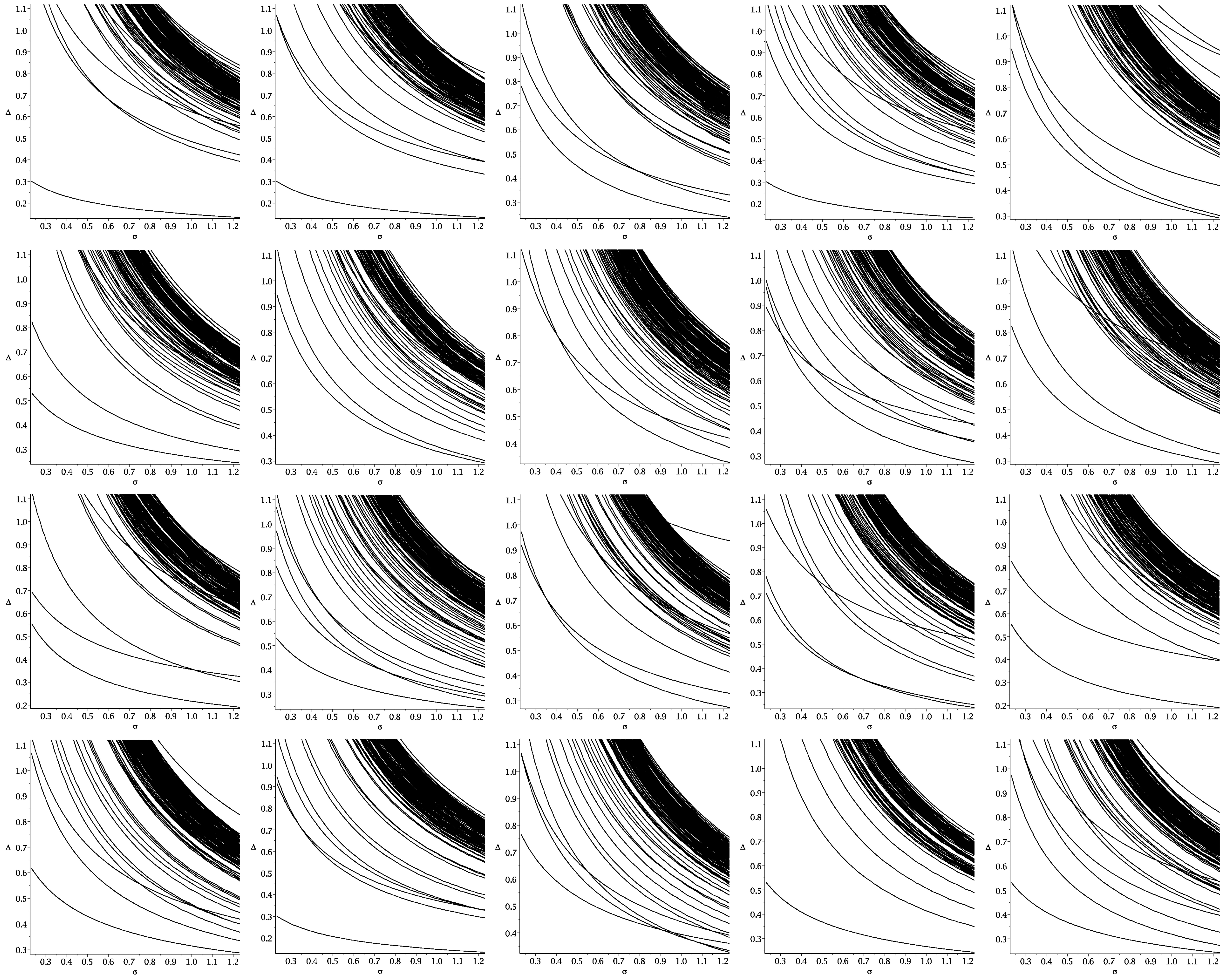}
 \caption{The level curves of the probability occurrences of amino acids for the Pfam protein family PF00060 from Clan CL00030.}
\end{figure}

The level curves corresponding to the values of $M_{j_s}\big(\sigma(a), \Delta(a)\big) = \frac{n_{J_s}(a)}{100}$ (the height of
the planes) which intersect the surfaces $p_j\big(\sigma(a), \Delta(a)\big)$ should be written for each $A_{kl}$ cell as:
\begin{align}
 &M_{j_{s}kl}\big(\sigma(a), \Delta(a)\big) = \nonumber \\
 &\resizebox{.98\hsize}{!}{$\frac{1}{\sigma(a)-l\big(\sigma_{n_{\min}}(a)-\sigma_{1_{\max}}(a)\big)}\Bigg(1- e^{j\big(\sigma(a)
 -l\left(\sigma_{n_{\min}}(a)-\sigma_{1_{\max}}(a)\right)\big)\big(\Delta(a)-(3-k)\left(\delta_{1_{\max}}(a)-\Delta_{\min}(a)
 \right)\big)}$} \nonumber \\
 &\resizebox{.92\hsize}{!}{$\cdot \sum\limits_{m=0}^{j}\frac{\left(j\big(\sigma(a)-l\big(\sigma_{n_{\min}}(a)-\sigma_{1_{\max}}(a)
 \big)\big)\big(\Delta(a)-(3-k)\big(\delta_{1_{\max}}(a)-\Delta_{\min}(a)\big)\big)\right)^m}{m!}\Bigg)$}
\end{align}
where we have used eq.$(34)$.

\section{Concluding Remarks}
In the present work we have introduced the idea of protein family formation process (PFFP) and we now stress that the evolution of an
``orphan'' protein is not a special case of this process \cite{mondaini3}. We then consider that proteins do not evolute
independently. Their evolution is a collective evolution of all their ``relatives'' grouped into families. In order to derive a model
of probability occurrence of amino acids, we have started from a master equation and we have made a very simple assumption such as
that of eq.$(25)$. We expected to obtain a good method for identifying the Clan association of protein families \cite{mondaini2} in
terms of level curves of the probability distributions of amino acids. The examples given into section 6, eqs.$(11)$, $(12)$, $(13)$,
in despite of their advantage over any conclusion derived from the usual analysis of histograms of figs.8, 9, 10, do not seem to lead
to an unequivocal conclusion about Clan formation. We then propose the following research lines for future development:
\begin{enumerate}
 \item The consideration of other families and clans into the analysis reported in section 6.
 \item The introduction of other assumptions for deriving the probability distributions. The Saddle Point Approximation seems to be
 a most convenient one.
 \item The consideration of models of distributions based on the joint probabilities of section 2.
 \item The introduction of generalized Entropy Measures as the selected functions of random variables instead of studying the
 elementary probabilities distributions of section 2.
\end{enumerate}

Some reports on recent results related to these proposals are now in preparations and will be published elsewhere.

\end{document}